\documentclass[twocolumn,showpacs,preprintnumbers,amsmath,amssymb,prb,
	superscriptaddress]{revtex4}
\usepackage[dvips]{graphicx}
\usepackage{bm}
\begin{document}
\title{Conductance and polarization in quantum junctions}
\author{P. Bokes } \email{pb20@york.ac.uk}
\affiliation{Department of Physics, University of York, Heslington, York
         YO10 5DD, United Kingdom}
\affiliation{Department of Physics, Faculty of Electrical Engineering and
        Information Technology, Slovak Technical University, Ilkovi\v{c}ova 3,
        812 19 Bratislava, Slovak Republic}
\author{R. W. Godby}
\affiliation{Department of Physics, University of York, Heslington, York
         YO10 5DD, United Kingdom}

\date{\today{}}

\begin{abstract}
We revisit the expression for the conductance of a general nanostructure 
-- such as a quantum point contact --
as obtained from the linear response theory. 
We show that the conductance represents the strength of the Drude singularity in 
the conductivity $\sigma(k,k';i\omega \rightarrow 0)$. Using the equation 
of continuity for electric charge we obtain a formula for conductance in terms 
of polarization of the system.  This identification can be used for direct 
calculation of the conductance for systems of interest even at the {\it ab-initio}
level. In particular, we show that one can evaluate the conductance from 
calculations for a finite system without the need for special ``transport'' boundary conditions. \\
\end{abstract}

\maketitle

\section{Introduction}
The conductance gives the current as a linear response to an applied voltage for 
a given finite sample, specified by its atomic geometry. Typically,
the sample of interest may be a thin chain of metallic atoms or a single 
molecule attached to metallic electrodes. The usual theory 
for {\it ab-initio} prediction of conductances for these nanoscale 
devices is based on the Landauer formula~\cite{Landauer57} 
\begin{equation}
	G = \frac{2 e^2}{h} T(e_{F}), \label{landauer}
\end{equation}
which identifies the conductance with the transmission probability for an electron 
at the Fermi level, $T(e_{F})$, to penetrate the sample.

Given the impressive success of the Landauer-B\"{u}ttiker 
equations~\cite{Buttiker85} in many mesoscopic phenomena, many authors tried 
to relate it to basic principles. Since we are concerned with linear 
response, the general Kubo linear response 
formalism~\cite{Kubo59} is well suited to this purpose. Among the first to do so were Economou and 
Soukoulis~\cite{Economou81} who presented a derivation for noninteracting 
electrons. This was subsequently vigorously 
discussed~\cite{Thouless81,Langreth81,Stone88}, since perfect 
transmission $T=1$ results in a finite conductance, not infinite as 
might be expected. Eventually the debate was reconciled with the realization that 
that Formula (\ref{landauer}) gives the conductance between {\it reservoirs} and 
$\frac{2e^2}{h}$ represents an unavoidable minimum contact resistance 
if such macroscopic electrodes are connected through a quasi-1D contact. 
However, there are no such electrodes in the Kubo-like derivations,
as has been pointed out by Landauer~\cite{Landauer90}, and so the result 
remained somewhat puzzling. The argument in favour of a Kubo-like derivation
was an introduction of an auxiliary external field acting in the leads
that compensates exactly the charge-density oscillations produced by the
homogeneous field in the sample, resulting in the ``4-point'' conductance
formula~\cite{Langreth81,Thouless81}
\begin{equation}
	G^{4P} = \frac{2 e^2}{h} \frac{T(e_{F})}{1-T(e_{F})} \label{landauer4p}
\end{equation}
that relates the current to the local drop in total electrostatic potential.
However, in many experiments and situations it is the 2-point conductance $G$ that is
required, in which case
Formula (\ref{landauer}) is frequently used 
for conductance calculations (e.g. for nanowires and metal-molecule-metal 
junctions).  Therefore, clarification of the relationship of this formula to the Kubo approach
remains necessary, and may lead to practical advances for {\it ab-initio} calculations.

Much of the confusion arising from the ``boundary and bulk contribution to 
the conductivity'' introduced by the ideal leads has been clearly resolved
by Sols~\cite{Sols91} who emphasized the importance of global charge 
conservation and its relation to gauge transformations. In that paper, however,
only the result (\ref{landauer}) was derived and discussed and the induced
charge-density oscillations were ignored.

The fundamental understanding of conductance as quantum-mechanical transmission 
relies on the picture of distinct electro-chemical potentials
for the two electrodes. These dictate the occupation of states in the non-equilibrium
sample. An alternative point of view (equally valid but much closer to the 
Kubo formalism) is that of electrons accelerated in the 
applied external field. The ability of the field to accelerate electrons
in extended or bulk materials has been used by Kohn~\cite{Kohn64} 
to distinguish between conductors and insulators where conductors 
are characterized by
$
	\lim_{\omega \rightarrow 0} \Im \{ \sigma(\omega,q=0)\} 
	\sim A/\omega, A \neq 0,
$
and the case $A=0$ characterizes insulators. This point of view, but applied to 
conductance of finite samples or mesoscopic conductors, has been considered 
by Fenton~\cite{Fenton92} who strictly differentiated between
a localized external field leading -- through acceleration of electrons over 
a finite distance -- to a finite conductance and yielding the Landauer 
formula, and a homogeneous field giving a response characteristic of the bulk 
material with $\omega^{-1}$ 
singularity. An approach based on coupled self-consistent 
equations for the current density, electron density and induced field,
implicitly also representing the ``conductance based on acceleration''
point of view, was set out in detail 
by Kamenev and Kohn~\cite{Kamenev01}. They focus on 
the 4-point conductance, which they define as the current divided
by {\it induced} potential drop, thereby completely neglecting 
the electrostatic drop in the external potential. This is appropriate in some cases,
since a finite constant external field over infinitely long 
leads contributes negligibly to the total potential drop around the sample.
It is important to note that their self-consistent field is obtained 
from the 1D Poisson equation, i.e. a system resembling parallel
planes of charge $e$ moving along their normal rather then electrons 
confined to a 1D wire. The former systems has far stronger screening
behaviour then the latter. More detailed calculations 
by Sablikov~\cite{Sablikov98}, who also used the Kubo formalism, show that 
the effective interaction between electrons in a 1D wire is not strong enough 
to screen the charge and a uniform current coexists with an induced dipole 
moment of the electron density in the wire.

In this paper we derive a formula for the steady-state conductance in terms of the 
irreducible polarization of the system of interest. Our starting point 
is application of an external field and following the time 
evolution for large times. We will show that only the $q=0$ Fourier component
of the field influences the steady-state current because the nonlocal
conductivity is a sharply peaked function of $q$, eventually becoming 
a Dirac delta function. {\it The weight of this
delta function has directly the meaning of conductance.} At the same time,
this singular behaviour also corresponds to the Drude singularity 
$\Im\{ \sigma(\omega)\}  \sim \omega^{-1}$ 
if we set $q=0$ and then take the limit 
$\omega \rightarrow 0$. It is interesting to note that from this point
of view we can identify the conductance with the Drude weight, a connection encountered 
in somewhat different circumstances in the conductive properties 
of extended systems~\cite{Castella95}. In addition to providing an interpretation 
of the conductance, our formula is suitable for practical 
{\it ab-initio} calculations. Using the continuity equation 
we circumvent the need for matrix elements of current operator, which 
are in general difficult to evaluate numerically, and express the conductance 
in terms of the irreducible polarization at small imaginary 
frequency -- a quantity that can be evaluated efficiently even using 
{\it ab-initio} methods that describe elecronic correlation~\cite{GW,GW1}. 

\section{Conductance in imaginary frequency}

Consider a system, infinitely long along the $x$ direction and finite, infinite
or periodic along the other two perpendicular directions. Along the $x$ axis
we apply an infinitesimally weak external electric field. The response of 
the system will be in general non-local in time and space, and quantitatively 
described by the Kubo linear response theory~\cite{Kubo59}:
\begin{eqnarray} \label{j-E}
	{\bf j}({\bf r},t) &=& \int_{-\infty}^{t} 
	\tensor \sigma({\bf r},{\bf r}';t-t') \cdot {\bf E}({\bf r}',t')d^3 r' dt', \\
	\tensor \sigma({\bf r},{\bf r}';t) &=& i \Theta(t) 
		\int_{0}^{-i\beta} d \tau \label{sigma-g}
                \mathrm{Tr}\left\{ \rho_{eq}  {\bf j}({\bf r}',t + \tau) 
		{\bf j}({\bf r},0) \right\},~
\end{eqnarray}
where ${\bf j}({\bf r},t)$ is the current density, ${\bf E}({\bf r},t)$ 
external electric field, $\tensor \sigma({\bf r},{\bf r}';t)$ non-local tensor 
of conductivity, $\Theta(t)=1$ for $t>0$ and $\Theta(t)=0$ otherwise 
and $\rho_{eq}$ the equilibrium density matrix. 

We take the form of the field as $E_{x}({\bf r},t)=E(x)\Theta(t), 
E_{y}=E_{z}=0$.  Introducing the current $I(x,t)=\int d{\bf S} \cdot 
{\bf j}({\bf r},t) $ (where we integrate over the surface perpendicular 
to $x$) and utilising our particular choice of external field we obtain
\begin{eqnarray}
	I(x,t) &=& \int dx' I_{\sigma}(x,x';t) E_{x}(x') \label{I-E} \\
	I_{\sigma}(x,x';t) &=& \int_{-\infty}^{t} dt'  
	\int d{\bf S} \cdot \tensor \sigma({\bf r},{\bf r}';t') \cdot d{\bf S}' 
	\label{I-sigma}
\end{eqnarray}
In the steady state we are interested in the limit $t \rightarrow \infty$
so that for the steady-state current we have
\begin{eqnarray}
        I_{\sigma}(x,x') &=& \lim_{\alpha \rightarrow 0^{+}} 
	\int_{-\infty}^{\infty} dt' e^{-\alpha t'} \sigma(x,x';t') \nonumber \\
	&=& \sigma(x,x';\omega=i0^{+}),
\end{eqnarray}
where we have introduced the effective one-dimensional conductivity $\sigma(x,x';\omega) = 
\int d{\bf S} \cdot \tensor \sigma({\bf r},{\bf r}';\omega) \cdot d{\bf S}'$. 
The positive infinitesimal $\alpha$ plays an important role for 
systems without dissipation, where application of a field for an infinitely
long time results in the system being heated. This is avoided by introducing 
an effective finite (but large) measurement time $T\sim \alpha^{-1}$.
In this way we avoid the heating problem because we first let 
$E(x) \rightarrow 0$ (linear response) and only afterwards 
$\alpha \rightarrow 0^{+}$. (Identical expressions are obtained assuming
adiabatic switching-on of the external field $E(x,t)=E(x)e^{\alpha t}$ as
$t$ approaches 0 from below, and measuring the current at $t=0^{-}$, which is the more 
conventional point of view.)

We are principally concerned with the total current $I$ and not the current 
density. Similarly, we would like to work with the bias voltage 
$V=\int d{\bf r} \cdot {\bf E}({\bf r})= \int dx E(x)$
rather than the field itself. To achieve this we Fourier-transform
with respect to $x$ 
\begin{eqnarray} 
	I(q) &=& \int \frac{dq'}{2 \pi} \sigma(q,q') E(q') \label{I-q} \\
	\sigma(q,q') &=& \int e^{-iqx+iq'x'} \sigma(x,x') dx dx'
\end{eqnarray}
First, the steady-state current needs to be independent of $x$, $I(x)=I$, as 
a consequence of the equation of continuity, so that $I(q) = 2\pi I \delta(q)$ 
and therefore directly also $\sigma(q,q') \sim \delta(q)$. 
Second, as a direct consequence~\cite{Sols91} of the linearity of the theory,
the steady-state current is uniquely given by the bias $V=E(q=0)$ only.
This means that for two external fields $E(q)$ and $E'(q)$ whose long-range parts are equal 
($E(q=0)=E'(q=0)$), but otherwise are arbitrary, one has to obtain identical 
steady-state currents. We therefore conclude that $\sigma(q,q') \sim 
\delta(q')$ in the limit $\alpha \rightarrow 0^{+}$, i.e. it is a sharply 
peaked function of $q'$ so that, based on Eq.~(\ref{I-q}),  it is only 
the values of $E(q)$ or $E'(q)$ at $q=0$ that affect the steady current. 
Using these observations in Eq.~(\ref{I-q}) we obtain
\begin{equation}
	G=\frac{I}{V}= \int \frac{dq dq'}{4 \pi^2} \sigma(q,q') \label{G-int}
\end{equation}
or using the fact that $\sigma(q,q') \sim \delta(q) \delta(q')$ we can also 
write 
\begin{equation} 
	\lim_{\alpha \rightarrow 0^{+}} \sigma(q,q';i\alpha) 
	= 4 \pi^2 G \delta(q) \delta(q'). \label{G-delta}
\end{equation}
This is one of the central results of this paper and directly shows that
conductance is the strength or weight of the Drude singularity. Formula
(\ref{G-int}) can be used for extrapolative evaluation of the conductance
from the conductivity at small imaginary frequencies, and since it is 
peaked function in $q,q'$, the conductivity needs to be evaluated for small
values of these variables only. We will demonstrate this
approach numerically in Section \ref{sec-4}.

\section{Conductivity and polarization} 
\label{sec-2}
Evaluation of the conductance using Equation (\ref{G-int}) requires
calculation of the Fourier transform of the conductivity. The latter
is, in a real-space representation, given by Equation (\ref{sigma-g}) which itself
is difficult to evaluate for all necessary ${\bf r},{\bf r}'$. On the other 
hand, substantial experience has been accumulated in calculations
of the polarization $\chi({\bf r},{\bf r}';t)$ , defined by~\cite{GW,GW1}
\begin{equation}
	\chi({\bf r},{\bf r}';t-t') = \frac{ \delta n({\bf r},t) }{\delta 
	V({\bf r}',t') },  \label{chi-def}
\end{equation}
where $\delta n({\bf r},t)$ is a change in density due to infinitesimally 
weak external potential $V({\bf r},t)$. To relate these two we utilize 
the equation of continuity integrated over the cross-sectional area
\begin{eqnarray}
	\partial_{x} I(x,t) &=& \partial_{t} N(x,t).
\end{eqnarray}
From the definition of the polarization we have 
$N(x,t) = \int_{0}^{t} dt' \int dx' \chi(x,x';t-t') V(x,t') $,
where our $\chi(x,x';t)$ is now, similarly to $\sigma(x,x';t)$, integrated 
across the cross-sectional area ($d{\bf S}$ and $d{\bf S}'$ integrals in 
(\ref{I-sigma})).
Substituting the linear response formulae for current (\ref{I-E}) and density
$N(x,t)$, Fourier-transforming into $q,q'$ variables and using the fact
that the external potential is arbitrary, we immediately obtain
\begin{equation}
        \sigma(q,q';i\alpha\rightarrow i0^{+}) = \lim_{\alpha \rightarrow 0^{+}}
        \frac{\alpha}{qq'} \chi(q,q';i\alpha). \label{sigma-chi}
\end{equation}
We note that the singular character of $\sigma(q,q')$ for small $q,q'$ 
as given in (\ref{G-delta}) does not arise fundamentally from the $1/qq'$ prefactor, since 
$\chi(q,q';\omega) \approx qq' f(q,q';\omega)$ where $f(0,0)\neq0$. 
This property is a consequence of conservation of total number
of particles ($n(q=0)=N$) or the absence of response in density if 
we change the potential everywhere by a constant $q'=0$.

Expressing the conductance through the polarization is particularly suited 
to a correct treatment of a system of interacting electrons. In the first place,
it is crucial to define the conductance not as a coefficient for current
dependence on the external but on the total electric field 
${\bf E}^{t}({\bf r}) = {\bf E}({\bf r}) + {\bf E}^{i}({\bf r})$. The induced 
field ${\bf E}^{i}({\bf r})$ can be obtained from Poisson equation 
\begin{equation}
	i (q \hat{{\bf x}} + {\bf G} ) \cdot {\bf E}^{i}(q,{\bf G};t) = 
	- 4 \pi \delta n(q,{\bf G};t),
\end{equation}
where $\hat{{\bf x}}$ is a unit vector in the $x$ direction and ${\bf G}$ 
is a 2D reciprocal lattice vector corresponding to the perpendicular
coordinates. Using the linear response result
for $\delta n(q,{\bf G};t)$ in terms of $\chi({\bf r},t)$ 
(see Eq.~(\ref{chi-def})) we arrive at the relation between the total 
and external field
\begin{eqnarray} 
	{\bf E}^{t}(q,{\bf G}) &=&  \int dq' \label{Et-E}
        \left[ \hat{\bf x} \delta_{{\bf G},0} \delta(q-q') \right. \\ & &
        + \left. \frac{q\hat{\bf x}+{\bf G}}{q'}\frac{4\pi}{q^2 + |{\bf G}|^2}
	\chi(q,\bm{G};q',0) \right] E(q') \nonumber 
\end{eqnarray}
If we multiply the latter equation by the unit vector $\hat{\bf x}$ and 
subsequently invert it for ${\bf G}={\bf 0}$, we obtain the relation
between the external field $E(q)$ and the total field along $\hat{\bf x}$ 
averaged over the cross-sectional area, $E^{t}(q,{\bf G}=0)$. 
Substituting this relation into equation (\ref{I-q}) we obtain, after some 
algebra (Appendix A),
\begin{equation}
	G = \lim_{\alpha \rightarrow 0^{+}} \frac{\alpha}{4 \pi^2} \int
	\frac{\chi^{t}(q,q';i\alpha)}{qq'} dq dq' , \label{G-chi-IR}
\end{equation}
where we introduce the ``{\it transport part of the polarization}'', $\chi^{t}(q,q';i\alpha)$, 
related to the irreducible polarization $\chi^{0}(q,{\bf G};q',{\bf G}')$
through an equation of Dyson type
\begin{eqnarray}
	\label{chi-dyson}
        \chi^t(q,{\bf G};q',{\bf G}')&& = \chi^{0}(q,{\bf G};q',{\bf G}') 
                \\ + \int dk \sum_{{\bf K} \neq {\bf 0}} && \chi^t(q,{\bf G};
		k,{\bf K}) \frac{4 \pi}{k^2 + |{\bf K}|^2} 
		\chi^{0}(k,{\bf K};q',{\bf G}'). \nonumber
\end{eqnarray}
The corrections entering through ${\bf K}\neq 0$ terms in 
equation (\ref{chi-dyson}) are known as local-field 
effects in the context of evaluation of the macroscopic dielectric 
function~\cite{GW1}. Here, however, it is not quite the same since 
$k=0, {\bf K\neq 0}$ part is included into $\chi^t$ whereas for 
the macroscopic dielectric function the sum is restricted to 
$k\neq0, {\bf K\neq0}$. The omission of the $\bf K = 0$ term 
involving $\frac{4\pi}{k^2}$ 
in equation (\ref{chi-dyson}) stems directly from the fact that 
it is precisely this term that converts the drop in external field 
into drop in the total field, as can be seen from Eq.~(\ref{Et-E})
when taking into account that the drop in total field is given by 
$\Delta V^t = E^t(q=0,{\bf G}=0)$.
Correct evaluation of the conductance in 3D therefore requires inclusion 
of these ``perpendicular'' local-field effects, included within $\chi^{t}$ 
but not in $\chi^{0}$, into account.  
Essentially, $\chi^{t}$ describes the response of the density to the effective 
potential, except that long-range screening of the potential in 
the $x$ direction is specifically excluded, allowing the conductance 
to address the applied voltage rather than the local potential drop.

For systems translationally 
invariant along the perpendicular directions, $\chi^{0}(q,{\bf 0};k,{\bf G}) \sim 
\delta_{{\bf 0},{\bf G}}$ and the last term in (\ref{chi-dyson}) becomes 
identically zero. It follows that in this particular case $\chi^t=\chi^{0}$ 
and the conductance of non-interacting electrons, defined with respect 
to drop in the {\it external} field, is identical to the conductance of interacting 
electrons, treated within the RPA approximation, defined with respect to drop 
in the {\it total} field.

\section{Landauer formula}
\label{sec-3}
	
In this section we present simple, analytically tractable cases
that illustrate the theory of the preceding sections. Consider first the case 
of non-interacting electrons in a quantum wire with only one subband. 
It is well known that the polarization function has the form~\cite{Williams74}
\begin{equation}
	\chi(q,q';\omega) =  
	\frac{1}{2\pi q} \log \left[ \frac{\omega^2 - (q^2/2 - k_{F}q)^2}{
                \omega^2 - (q^2/2 + k_{F}q)^2} \right] \label{chi-omega}
	\times 2 \pi \delta(q-q'),
\end{equation}
where the factor $2\pi \delta(q-q')$ arises trivially from the translational
invariance of the system along the $x$ axis. 
Using this expression in (\ref{sigma-chi}) immediately gives $\sigma(q,q';i\alpha \rightarrow i0^{+}) = 
2 \pi \delta(q') \delta(q)$ and therefore through Eq.~(\ref{G-delta})
$G=1 / 2 \pi$, the quantum of conductance (i.e. $e^2/h$ in S.I. units).
Using Eq.~(\ref{G-chi-IR}) without application of the limit, and 
Eq.~(\ref{chi-omega}), it is also possible to obtain the analytical dependence 
of conductance $G$ on imaginary frequency $\alpha=-i\omega$ for this system,
\begin{equation} 
	G(\alpha) = \label{G-alpha-ana}
		\frac{1}{\sqrt{2}\pi\sqrt{1 + \sqrt{1+[\alpha/e_{F}]^2} } } .
\end{equation}
This functional form will be extremely useful for numerical calculation
of $G(0^+)$ based on extrapolation to zero frequency, described in 
Section~\ref{sec-4}.

It is also instructive to explore the behaviour of 
$I_{\sigma}(t \rightarrow \infty)$ directly (see Eq.~(\ref{I-sigma})). 
The latter can be calculated, e.g. using
Fourier transform of (\ref{chi-omega})
\begin{eqnarray}
	I_{\sigma}(q,q';t) &=& \int_{0}^{t} dt' \sigma(q,q';t') 
	   = \int_{0}^{t} dt' \frac{ \partial_{t'} \chi(q,q';t') }{qq'} 
		\nonumber\\
	&=& \frac{ \chi(q,t) }{q^2} \times 2 \pi \delta(q-q'),
\end{eqnarray}
where we have used the expression (\ref{sigma-chi}) in the real-time domain.
The result is
\begin{equation}
	I_{\sigma}(q,q';t) = \label{I-sigma-0}
        \frac{2}{\pi q^3 t} \sin(k_{f}qt) \sin(\frac{q^2}{2}t) \Theta(t)
	\times 2 \pi \delta(q-q'),
\end{equation}
which for $t \rightarrow \infty$ evidently approaches the 
form (\ref{G-delta}) with $G=\frac{1}{2\pi}$. The lesson from these analytical
examples is that the well-known formulae for polarization can be used for
evaluation of the conductance of the system, and that the limit $i\omega 
\rightarrow 0^+$ clearly corresponds to the steady-state limit $t\rightarrow 
\infty$. 

Because the conductivity is sharply peaked around $q'=0$, only 
the $q=0$ component of the applied field $E$ has an effect on the total 
current. This allows us to include, as a limiting case, a homogeneous field. 
For a system with a constant field applied over
length $L$, the field itself must be $\sim 1/L$ so that we have
a sufficiently small finite drop $V = \int_{L} E dx \sim 1$. The physical 
meaning of the finite conductance of a scattering-free segment of a metallic 
wire is clearly a manifestation of the free acceleration of electrons over 
the distance $L$. This is in agreement with the point of view advocated 
by Fenton~\cite{Fenton92}. 
However, there is no need to keep $L$ strictly finite; in fact 
the limiting case $L\rightarrow \infty$ can be still characterized by a finite 
overall conductance by taking careful limits and using properties of the Dirac delta function, as mentioned above for the case of the free electron gas 
(following Eq.~(\ref{I-sigma-0})).
In simple terms, this represents the limit of increasing the length over which 
the external field is applied, $L$, while decreasing the intensity of the 
field $E$ in such a way as to keep the drop $V \sim E L$ finite and small.

\begin{figure}[t]
\begin{center}
\includegraphics[width=8cm]{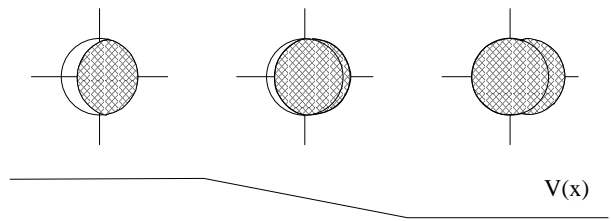}
\caption{Semi-classical interpretation of a finite-field potential $V(x)$ 
applied to a wire. The local Fermi spheres are depicted as for 2D gas for 
clarity. The filled area on top of the equilibrium spheres indicates 
accelerated electrons, while the empty area inside the equilibrium spheres 
represents states emptied by the deccelerating effect of the  
uphill field.} \label{fig-1}
\end{center}
\end{figure}

A simple qualitative demonstration of this argument can be made 
semi-classically. Consider a finite drop over finite region of a 1D wire 
as shown in Fig.~\ref{fig-1}.
The current can be obtained from the local Fermi distribution on the far right.
Electrons that are occupying states above the equilibrium distribution 
have travelled to the right from the region with an accelerating field. These, during their flight
through the region, gained energy $\Delta V$ so that they represent current
$I = \frac{1}{L} \sum_{k} k_{F} $ where we sum over states with accelerated 
electrons, i.e. $k: e_{k} \in (e_{F},e_{F}+\Delta V)$. This obviously leads
to the quantum of conductance $G=\frac{1}{2 \pi}$. While formally we have 
striking similarity with the usual "two chemical potentials" picture, 
it should be noted that the microscopic interpretation of these expressions 
is different. As can be seen from Fig.~\ref{fig-1}, the local charge 
neutrality is clearly violated in the constant potential regions on the right 
and left while in the center it is locally charge neutral where 
the distribution corresponds to the shifted Fermi sphere. The infinite $L$ 
limit, discussed in previous paragraph, resolves this problem by sending 
the regions with unbalanced local charge away into infinities. 

When we insert a localized scattering potential characterized by the 
transmission matrix 
\begin{equation}
	{\bf T}(k) = \left[ \begin{array}{cc}
		r(k) & \tilde{t}(k) \\
		t(k) & \tilde{r}(k) 
		\end{array} \right]
\end{equation}
into 1D gas of electrons, the Landauer formula is obtained. We obtain 
the desired demonstration of the sharply-peaked character of $\sigma(q,q';i\alpha)$ 
by analytically continuing Kamenev and Kohn's expression~\cite{Kamenev01} 
onto the imaginary frequency axis, 
\begin{eqnarray}
	\sigma(q,q';i\alpha) &=& \frac{1}{2 \pi} \left\{
		\frac{2 k_{F} \alpha}{k_{F}^2 q^2 + \alpha^2} 2 \pi \delta(q-q')
		- |r(e_{F})|^2 \right. \nonumber\\
		& &\times \left. \frac{ 2 k_{F} \alpha}{k_{F}^2 q^2 + \alpha^2}
		\frac{ 2 k_{F} \alpha}{k_{F}^2 q'^2 + \alpha^2} \right\} 
		\\	
	&\rightarrow& 2 \pi |t(e_{F})|^2 \delta(q) \delta(q'). \label{Landauer}
\end{eqnarray}
Alternatively, we can arrive at the same result using Equation 
(\ref{G-chi-IR}), which explicitly shows the utility of reformulation 
of transport through the polarization function in imaginary frequency.
A detailed derivation is given in Appendix B. 

We stress that this result for the {\it two}-point conductance (\ref{landauer}) 
of non-interacting electrons also applies to interacting electrons if 
the conductance is defined with respect to the total field at the RPA level 
of approximation.
The 4-point conductance (\ref{landauer4p}) may be formulated in a similar 
fashion, which, if combined with the approximate effective electron-electron
interaction $v_{ee}^{1D}(q) \sim \frac{1}{q^2}$, would be equivalent to the
treatment of Kamenev and Kohn~\cite{Kamenev01}.(In their calculation 
the 4-point conductance is defined as $G^{4P} = I/V^{i}$, where $V^{i}$ is the 
self-consistent induced drop only. The reason for neglecting the drop
in external field is due to the above mentioned limiting procedure;
since $E\times L$ is finite while $L\rightarrow \infty$, $E \times L'$ will 
be negligible for any finite $L'$. Since the 4-point measurement is meant 
across a finite distance $L'$, the contribution to the total drop on finite 
distance comes solely from the induced field.)

\section{Feasibility for numerical calculations}
\label{sec-4}

The expression for the conductance (\ref{G-chi-IR}), together 
with some experience gained from the evaluation of $G$ for 
lead-scattering-lead system in the Appendix, motivates the following 
suggestion: can the correct conductance of an infinite system be
calculated using a finite system with some arbitrary
boundary conditions, determining the polarization as a function of $\alpha$, and 
then extrapolating $\alpha \rightarrow 0$?

For the purpose of this exploration we have considered a square-barrier 
potential in 1D, for which the results for transmission coefficients 
are well known analytically. We calculate the polarization function using 
the Green's function of the equilibrium system given by equation
\begin{equation}
	\left[ e-\hat{H}(x) \right] \mathcal{G}(x,x';e) 
	= \delta(x-x'), \label{Green-eq}
\end{equation}
supplemented by the chosen boundary conditions. For simplicity and for 
the purpose of illustration, we take the extreme case of ``zero'' boundary 
conditions on the wavefunctions~\cite{Matsuoka03}. In this case, it is clear 
that $G(\alpha) \rightarrow 0$ as $\alpha \rightarrow 0$. 
In terms of the Green's function we can easily express the polarization as 
\begin{eqnarray}
	\chi(x,x';i\alpha) =&& 
	\sum_{i}^{occ} 
	\left( \mathcal{G}(x,x';e_{i}+i\alpha) \phi_{i}(x') \phi_{i}^{*}(x) 
	\right.  \nonumber\\
	&& +\left. \mathcal{G}(x',x;e_{i}-i\alpha) \phi_{i}(x) 
	\phi_{i}^{*}( x') \right), \label{chi-G-tim}
\end{eqnarray}
where we sum over occupied states only, and $\phi_{i}$ are
eigenstates of $\hat{H}$ from (\ref{Green-eq}). The Green's function
at given energy is found by direct integration of (\ref{Green-eq}) 
and eigenstates are easily found by matching plane waves with the chosen
boundary conditions. Finally we use a discrete Fourier transform,
$\chi(x,x') \rightarrow \chi(K,K')$, to obtain the estimate of conductance
\begin{equation}
	G(\alpha) = \label{G-num}
		 \sum_{K,K'} \alpha \frac { \chi(K,K';i\alpha)}{KK'}.
\end{equation}
The discrete Fourier transform introduces a convergence parameter: the 
real-space step size $\Delta x$ or, equivalently, the number 
of discretization points $N$. Altogether, the calculation needs to 
be converged with respect to the system size $L$ and with respect to $N$, 
and extrapolated with respect to imaginary energy $\alpha$. 

\begin{figure}[t]
\begin{center}
\includegraphics[width=8cm]{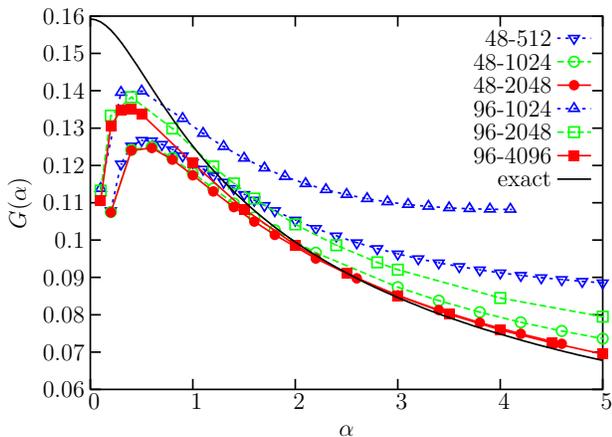}
\caption{The dependence of conductance of free electron 1D gas 
on imaginary frequency. Black line corresponds to the analytical result
(\ref{G-alpha-ana}) and the numerical data labels ``$L$-$N$'' represent 
system of length $L$ a.u. with $N$ discretization points in real space.
Apart from the low-energy drop to zero, which is characteristic of a finite 
system, the functions converge to the exact result for the infinite 
system.  This opens the possibility of calculating the D.C. conductance 
by extrapolation from moderate values of $\alpha$.} \label{fig-3}
\end{center}
\end{figure}

Before we discuss results for non-zero barrier height, let us first consider 
the case of zero barrier height, i.e. the free electron gas.
In this case we know exactly the whole dependence of $G(\alpha)$, given 
by Eq.~(\ref{G-alpha-ana}). We have fixed the Fermi wave-vector to $k_{F}=
\pi/3$~a.u., corresponding to $e_{F} \approx 0.5$~a.u. In figure \ref{fig-3}
we show numerical results obtained using Eq.~(\ref{G-num}). The labels 
``$L$-$N$'' represent system of length $L$ a.u. with $N$ discretization points 
in real space, or equivalently, the number of $K$ points in the discrete
Fourier transform. We obtain our chosen Fermi energy for 
lengths $L=48$ and $96$ when occupying $16$ and $32$ states respectively. 

From the figure (\ref{fig-3}) we see that the numerical results converge
to the analytical expression for energies $\alpha \gg e_{F}$. 
The lower limit 
of this range can be brought closer to $e_{F}$ by increasing 
the length of the system. However, for a longer system the convergence 
with respect to $N$ becomes more demanding, with the error growing with 
energy. This behaviour can be traced to the cusp in the polarization 
$\chi(x,x')$ at $x=x'$. It can be easily found from 
Eq.~(\ref{chi-G-tim}) that the size of this cusp is 
\begin{eqnarray}
        \chi'(x,x')|_{x=x'^{+}} - \chi'(x,x')|_{x=x'^{-}} =
        4 \sum_{i}^{occ} \phi^{*}_{i}(x) \phi_{i}(x).
\end{eqnarray}
This expression could be used to remove the cusp from numerical Fourier
transforms and substantially decrease the needed number of $K$ points.
In the present paper, since our system is computationally undemanding, we have 
instead simply increased $N$ until satisfactory convergence was achieved. 

\begin{figure}[t]
\begin{center}
\includegraphics[width=8cm]{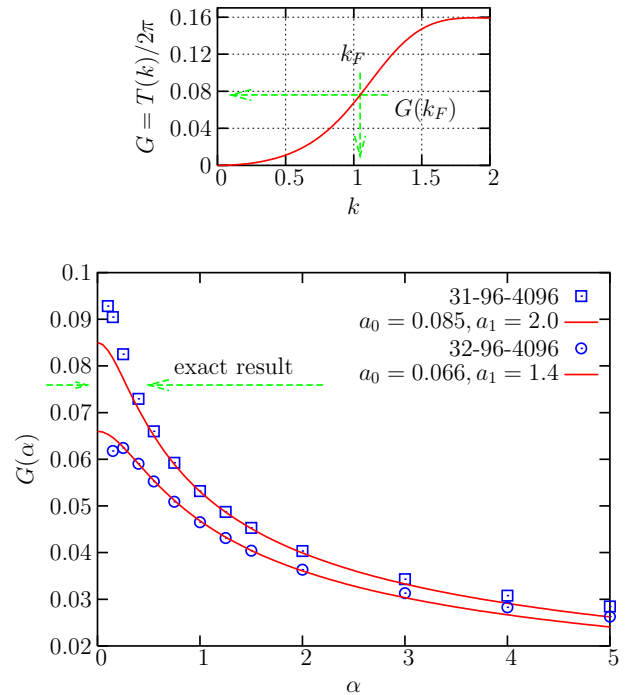}
\caption{The dependence of conductance, given by Landauer formula, on
the Fermi energy ($k_{F}=\pi/3$ in our calculations) for a square barrier 
of height $e_{F}$ and width 
$\lambda_{F}/3$ (upper graph) and the dependence of the numerically 
calculated conductance on the imaginary energy (lower graph). 
$G(\alpha)$ approaches the exact result of infinite system before 
it turns rapidly towards zero.} \label{fig-2}
\end{center}
\end{figure}

Next we include a square barrier into our system. In Fig.~\ref{fig-2} 
we show results of calculation with barrier with width $a=2$ a.u. and 
height $e_{F}$. In this case the transmission probability is approximately 
50\% which is indicated in the upper graph with arrows. 
In the lower graph we show the dependence of conductivity on the imaginary
frequency $\alpha$. The two sets of numerical data correspond to odd 
and even numbers of occupied states, converged with respect to $N$ 
within the considered range of $\alpha$. The convergence of data with respect 
to $L$ is first fast but eventually becomes rather slow. In fact, 
it is not our aim to get the conductance of an ``almost infinite'' system, 
but rather to be able to extract the conductance of an infinite system from 
a calculation for a sufficiently large
but finite system. The shown length $L=96$ a.u. is to be compared with the Fermi 
wavelength $\lambda_{F} \approx 6$. 

A striking difference between odd and even number of occupied states comes 
from the fact that placing the barrier of length $a=2$ in the center affects 
the symmetric and antisymmetric states (at the Fermi energy) differently. 
Clearly, the former have a much larger amplitude at the scattering potential 
than the latter (which are zero in the center of the system) 
and therefore the symmetric states will be more affected. This is in agreement 
with the conductance for $32$ occupied states being below that for $31$ occupied 
states. For $L\rightarrow \infty$ these two states become degenerate and the
difference between odd and even cease to exist. By bracketing the exact conductance in this way, 
we can obtain acceptable convergence even with the crude ``zero'' boundary conditions.

When calculating conductance we need to be able to extrapolate our 
data to zero energy. As we have pointed out above,
the small-energy turnaround of $G$ arises from the finite size of the system
and we should not take that into account. To fit the data we have used 
a scaled version of the functional form obtained for the free gas, 
\begin{equation}
	G(\alpha) = 
	\frac{a_{0}}{\sqrt{2}\pi\sqrt{1 + \sqrt{1+[2 a_{1} \alpha]^2} } }, 
\end{equation}
where $a_{0}$ and $a_{1}$ are fitting parameters, representing linear 
scaling of the conductance and energy axes. As we see from Figure
\ref{fig-2} the fit is very good over the crucial middle
range of frequencies. For $\alpha > 3$ it becomes worse but we already
know that this difference can be attributed to convergence problems
with number of $K$ points due to the cusp in $\chi(x,x')$. When 
looking at $\alpha \rightarrow 0$ limit of our fitted expressions 
(coefficients $a_{0}$ in the figure \ref{fig-2}) we see the 
values bracket the exact value with an relative error of roughly 10\%.  
Averaging the result for even and odd states gives a smaller and more 
convergent error (0.5\%).

This calculation is intended to be illustrative, yet it shows
that the formula (\ref{G-chi-IR}) can, in principle, lead to the correct
answer even when applied to system with ``zero'' boundary conditions.
We particularly believe that the use of periodic boundary conditions
will significantly improve the performance of this approach for the main 
problem of the zero boundary condition is clearly the one-way transfer 
of charge from one side to the other, which introduces finite-system errors
after a relatively short time, and 
therefore spoils $G(\alpha)$ between $\alpha=0$ and a relatively large value.
This gross effect is not present for a periodic system. 
At the same time, periodic boundary conditions remove the unwanted artificial oscillation 
in $G$ with the number of occupied states, arising from the even/odd symmetry as 
discussed above.

\section{Conclusions}

We have developed a unifying point of view of the
polarization, the nonlocal conductivity and the conductance which supplies a steady-state 
transport characteristics of any system. We have shown that the weight
of the Drude singularity at zero frequency of the nonlocal conductivity, when 
considered in reciprocal space, directly corresponds to the 
conductance of the system to which we apply a field with a nonzero overall drop
in potential. We have identified a simple relation between
conductivity and polarization for the case of a system under unidirectional
external field that eventually led us to an simple formula for conductance,
expressed through the polarization of the system at small imaginary 
frequency. Expressed in terms of polarization, it turned out to be possible
to address the self-consistent field that contributes to the drop in total
potential used for definition of conductance. We have shown
that the formula for its evaluation remains formally intact but instead 
of the polarization function we need to supply a ``transport part'' of the polarization.
The latter is identical to the irreducible polarization in
1D, but differs from it in general 2D or 3D systems except where perfect translational invariance 
exists perpendicular to the flow of current.

Finally we have demonstrated that our expression for conductance in terms
of polarization can be used for the convenient numerical evaluation of the conductance 
for systems without imposing specific boundary conditions in the form 
of scattering states. This formulation 
is directly suitable for inclusion of many-body or inelastic effects, since
it is based on the polarization function for which approximations that 
include these complications are very well developed.

\begin{acknowledgments}
The authors gratefully acknowledge useful discussions with Angel Rubio. 
This work was supported by the RTN programme of the European Union NANOPHASE 
(contract HPRN-CT-2000-00167).
\end{acknowledgments}

\section*{Appendix}

\subsection{Derivation of equation (\ref{G-chi-IR})}
\label{app-a}

Let $g(q,q')$ be the inverse operator to the right hand side in (\ref{Et-E})
for ${\bf G=0}$.
\begin{equation}
	\int dk g(q,k) \left( \delta(k-q') -  \label{g}
	\frac{4 \pi}{kq'}\chi(k,q') \right)  = \delta(q-q') ,
\end{equation}
so that $E(q) = \int dk g(q,k) E^{t}(k)$. According to (\ref{I-E}) we have
\begin{equation}
        I(q) = \int dk \int \frac{dq'}{2 \pi} \label{I-E-app}
        \frac{ \alpha \chi(q,q')}{qq'} g(q',k) E^{t}(k),
\end{equation}
which motivates us to define 
\begin{equation}	\label{chi-t-def}
	\chi^{t}(q,q') = q' \int dk \frac{\chi(q,k)}{k}g(k,q') .
\end{equation}
Multiplying (\ref{g}) with $q' \chi(q'',q)/q$ from left and 
integrating over $q$ we have
\begin{equation}
	\chi^{t}(q'',q') - 
	4 \pi \int dk \frac{\chi^{t}(q'',k)}{k^2}  \chi(k,q')
	= \chi(q'',q'),
\end{equation}	
comparing term-by-term this Dyson-like equation with the Dyson equation
for the reducible polarization $\chi(q,q')$ we arrive at the equation
(\ref{chi-dyson}) which omits the ${\bf K}=0$ term from the sum. Substitution
of (\ref{chi-t-def}) into (\ref{I-E-app}) give immediately the result
(\ref{G-chi-IR}) which concludes the stated results in section \ref{sec-2}.
	
\subsection{Derivation of equation (\ref{Landauer}) using 
	formula (\ref{G-chi-IR})}
\label{app-b}

The polarization of non-interacting electrons is given by 
expression~\cite{GW}
\begin{equation}
	\chi(q,q';i\alpha) = \sum_{ij} \label{chi-sum}
	\langle i | e^{-iqx} | j \rangle \langle j | e^{iq'x} | i \rangle
                \frac{n_{i} - n_{j}}{i\alpha - e_{j} + e_{i}}, 
\end{equation}
where $|i\rangle$ are eigenstates of the electronic Hamiltonian with eigenenergy
$e_{i}$, $n_{i}$ is its occupation factor and 
\begin{equation}
	\langle i | e^{-iqx} | j \rangle =
	\int dx \langle i | x \rangle e^{-iqx} \langle x | j \rangle.
\end{equation}
Using formula (\ref{G-int}) we therefore have
\begin{equation} \label{G-int-chi0} 
	G = \int \frac{dq}{2\pi} \frac{dq'}{2\pi} \sum_{ij} 
	\langle i | \frac{e^{-iqx}}{q} | j \rangle 
	\langle j | \frac{e^{iq'x}}{q'} | i \rangle
                \frac{\alpha(n_{i} - n_{j})}{i\alpha - e_{j} + e_{i}}. 
\end{equation}
The first two matrix elements are, after integrations, complex 
conjugate to each other and therefore their product is real. 
The last fraction in (\ref{G-int-chi0}) is not state- but energy-dependent 
and its real part, which is only needed, is given by 
\begin{equation}
	\frac{\alpha}{(e_{i}-e_{j})^2 + \alpha^2} \frac{ dn( e_{i} ) }{d e_{i}}
	(e_{i}-e_{j})^2. \label{occ-fac}
\end{equation}
We have already used the fact that in the limit of our interest ($\alpha
\rightarrow 0$) the first factor will be sharply peaked and therefore
we can use linear Taylor expansion of $n(e_{j})$ around $e_{i}$. However,
the expression will be non-zero only if the factor $(e_{i}-e_{j})^2$ 
will be compensated by the energy dependence of the matrix elements,
which we will confirm in following paragraphs.

In the next step we interchange the order of integrations in (\ref{G-int-chi0})
so that we directly evaluate
\begin{equation}
	\label{theta} \int \frac{dq}{2 \pi} \frac{e^{iqx}}{q + i\epsilon} 
	= -i \Theta(-x) e^{\epsilon x} ,
\end{equation}
where we interpret the singularity to be at $q=-i\epsilon$. This is in fact
arbitrary since, as we have pointed out in the section \ref{sec-2},
$\chi(q,q';\omega) = qq'f(q,q')$. An equally valid choice of singularity 
at $q=+i\epsilon$ leads to identical results.

We can now finally turn to evaluation of the matrix elements. The asymptotic 
character of the wavefunctions has the well known form
\begin{eqnarray}
	\phi_{k,R}(x) &=& \frac{1}{\sqrt{2\pi}} \left\{ \begin{array}{ll}
		e^{ikx} + r_{k} e^{-ikx} & x \ll 0 \\
		t_{k} e^{ikx} & x \gg 0 
		\end{array} \right. \\
	\phi_{k,L}(x) &=& \frac{1}{\sqrt{2\pi}} \left\{ \begin{array}{ll}
		\tilde{t}_{k} e^{-ikx} & x \ll 0 \\
		 e^{-ikx} + \tilde{r}_{k} e^{ikx} & x \gg 0 
		\end{array} \right.
\end{eqnarray}

The final integrals we need to do have the form 
\begin{equation}
	( k, R/L | k', R/L ) =
	\int_{-\infty}^{0} e^{\epsilon x} \phi_{k,R/L}^{*}(x) \phi_{k',R/L}(x) 
	dx,
\end{equation}
and of there we need to keep only those that are singular $\sim \frac{1}{k-k'}$
since in (\ref{occ-fac}) we need to compensate the factor
$(e_{k}-e_{k'})^2 = \frac{1}{2}(k+k')^2(k-k')^2$. Foreseeing the delta-function
character with respect to $e_{k}-e_{k'}$ and $e_{k}-e_{F}$ appearing 
in (\ref{occ-fac}) we can directly use $t_{k}=t_{k'}=t_{k_{F}}$ and 
$r_{k}=r_{k'}=r_{k_{F}}$, the transmission and reflection probability 
amplitudes at the Fermi energy respectively. It is now easy to see that 
the singular terms are 
\begin{eqnarray}
	(k,R|k',R) &=& - (k,L|k',L) = \frac{1}{2\pi} \frac{|t|^2}{k-k'} \\
	(k,R|k',L) &=& (k,L|k',R)^{*} = 
				\frac{1}{2\pi} \frac{ r^* \tilde{t} }{k'-k} .
\end{eqnarray}
Using these forms, together with (\ref{occ-fac}) and the form zero temperature
limit that $dn(e)/de = \delta(e-e_{F})$ we directly get
\begin{equation}
	G = \frac{1}{2\pi} ( |t|^4 + |t|^2 |r|^2 ) = \frac{|t|^2}{2 \pi},
\end{equation}
the celebrated Landauer formula.

\end{document}